\documentclass[journal]{IEEEtran}

\usepackage{graphicx}
\usepackage{amsmath}

\usepackage[keeplastbox]{flushend}

\begin{document}




\title{Non-volatile Hierarchical Temporal Memory: Hardware for Spatial Pooling}
\author{	\IEEEauthorblockN{Lennard G. Streat$^{*}$, Dhireesha Kudithipudi$^{\dagger}$, Kevin Gomez$^{\S}$ }\\
 \IEEEauthorblockA{Nanocomputing Research Laboratory, Rochester Institute of Technology, Rochester, NY 14623$^{*\dagger}$\\
		Seagate Technology, Shakopee, MN 55379$^{\S}$\\
		Email: lgs8331@rit.edu$^{*}$, dxkeec@rit.edu$^\dagger$,kevin.gomez@seagate.com$^{\S}$}}



\maketitle

\begin{abstract}
Hierarchical Temporal Memory (HTM) is a biomimetic machine learning algorithm imbibing the structural and algorithmic properties of the neocortex. Two main functional components of HTM that enable spatio-temporal processing are the spatial pooler and temporal memory. In this research, we explore a scalable hardware realization of the spatial pooler closely coupled with the mathematical formulation of spatial pooler. This class of neuromorphic algorithms are advantageous in solving a subset of the future engineering problems by extracting non-intuitive patterns in complex data. The proposed architecture, Non-volatile HTM (NVHTM), leverages large-scale solid state flash memory to realize a optimal memory organization, area and power envelope. A behavioral model of NVHTM is evaluated against the MNIST dataset, yielding 91.98\% classification accuracy. A full custom layout is developed to validate the design in a TSMC 180nm process. The area and power profile of the spatial pooler are 30.538mm\textsuperscript{2} and 64.394mW,  respectively. This design is a proof-of-concept that storage processing is a viable platform for large scale HTM network models.
\end{abstract}


\section{Introduction}
\IEEEPARstart{H}{ierarchical} temporal memory (HTM) is a biomimetic machine learning algorithm, designed with the aim of capturing key functional properties of the mammalian brain's neocortex to solve pattern recognition problems. HTM theory was originally proposed by Jeff Hawkins in \cite{Hawkins2004}, and subsequently formalized in \cite{George2005, George2009}. Since its inception, HTM has evolved as a machine intelligence algorithm with its core being a cortical learning algorithm (CLA) \cite{Numenta2006, Numenta2007, Numenta2011}. Given  spatio-temporal data, HTM can use learned representations to perform a type of time-dependent regression\cite{Mnatzaganian2016}. Few applications that have shown promise in this area include predicting taxi passenger counts \cite{Cui2015}, and anomaly detection in streaming data.

\begin{figure}[t]
	\centering
	\includegraphics[width=\linewidth]{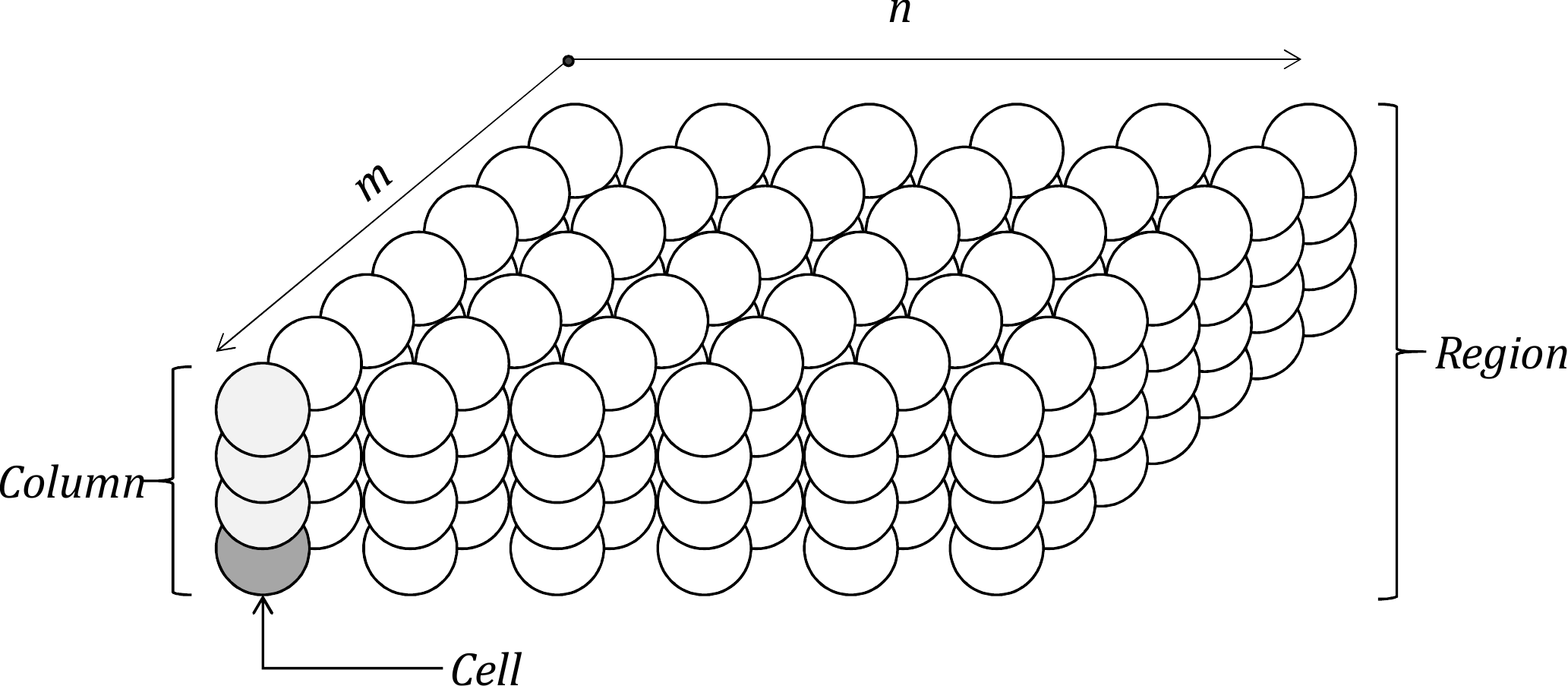}
	\caption[]{Example HTM region, consisting of a $6\times7$ column space ($n\times m$); each column is composed of 4 cells. Columnar computation is handled by the spatial pooler, and cellular-level operations occur within the temporal memory.}
	\label{fig:htm_region}
\end{figure}

In its most recent theoretical formulation, HTM consists of two main functional units: a spatial pooler(SP)  and a temporal memory (TM). These two substructures model the spatial and temporal patterns within data, respectively. The SP is responsible for mapping an input data on to a sparse distributed representation and the TM is responsible for 
learning sequences and making predictions. When combined, they form what is referred to as a ``\emph{region}''. 
An HTM region, depicted in Fig. \ref{fig:htm_region}, consists of multiple columns (SP). Each column consists of four cells (TM). These cells use proximal dendrites to connect the feedforward input, via synapses, to a column. There are distal segments which  connect cells within a region. Regions may be connected hierarchically to form larger systems. A region may be theoretically n-dimensional in size. For more structural details of HTM, please refer to \cite{Numenta2015}.

The focus of this research is on the SP, which maps encoded input (referred to as an ``\emph{input region}'') into a higher-dimensional space. Output generated by an SP region is an N-dimensional sparse binary vector or a sparse distributed representation (SDR)  \cite{Ahmad2015}. Spatial pooling is decomposed into three phases: namely, overlap, inhibition, and learning. Overlap computes the number of active-connected synapses on a column's proximal segment: the set of synapses mapping the input region onto the column. A synapse is considered to be connected if it's permanence (or weight) is above a fixed threshold and active if it is connected to an active input. Inhibition prunes the set of columns with proximal segments above the overlap threshold to obey the SDR constraint. Finally, learning rules are employed to enforce fairness amongst columns (boosting), and to train proximal segments to model patterns in the input region.

CLA was discussed in the literature, with emphasis placed on acceleration methods \cite{Zhou2012, Price2011}, noise resilience \cite{Mnatzaganian2016, Ahmad2015}, in addition to anomaly detection applications \cite{Chandola2009}. Despite these efforts, hardware implementations of CLA are still scarce: the same is true for Zeta \cite{Fan2014}. This is due in part to volatility of the algorithmic formulation. HTM CLA is studied, in part, in \cite{Numenta2011, Chen2012}. In this work emphasis is placed on studying the existing CLA models for hardware realization.

Following Numenta's official CLA white paper \cite{Numenta2011}, HTM has seen growth in research interest. 
For instance, a multi-core software-based CLA was presented in \cite{Price2011}. Zhou and Lou designed a C-based many-core implementation that leveraged the Adapteva hardware platform \cite{Zhou2012}. In the work of Liddiard et al., SP was implemented and evaluated using pseudo-sensory data \cite{Liddiard2013}. However, the analysis was lacking with regard to specificity in the methodology; a non-standard dataset was presented and not characterized in depth. Vyas and Zaveri presented a verilog implementation of SP \cite{Vyas2013}. More recently, Zyarah designed a 2D mesh architecture that modeled many concepts of CLA and resulted in significant speedup over a matlab-based software implementation \cite{Zyarah2015}. These architectures share some commonalities, and primarily focused on performance scaling, but discussions regarding hardware scalability are lacking. More recently, CLAASIC, a 16-core packet-switched 2D torus network was developed to model CLA \cite{Puente2016}. However, as with the previously mentioned models, online learning was not proposed.

In this research, we explore a nonvolatile memory-centric design for realizing the spatial pooler. As dynamic random access memory (DRAM) is reaching its scaling limits, semiconductor technology roadmap projections show that it will likely be replaced by other emerging nonvolatile memories such as spin-transfer torque magnetoresistive random access memory, resistive random access memory, or phase-change memory. Therefore, it is timely to investigate a nonvolatile memory centric processing approach for these systems \cite{prezioso2015training,hu2015associative,burr2015experimental}. Moreover in the future, it is expected that large information processing systems will move from message passing interface to memory-centric programming models, as network latencies improve. In the current market, a non-volatilve memory (NVM), such as flash, provides the optimal balance of cost, performance and energy for the data store in this data-centric architecture. To increase compute parallelism, concurrency (more parallel threads), and minimize data movement and co-locate HTM close to memory, we study flash within solid-state drives (SSD). 

In general, this research falls under a broader umbrella of neuromorphic computing. The thesis of modern neuromorphic computing is that if the key principles of brain functionality can be formulated and emulated in a hardware platform, we will be capable of solving a subset of the challenging future engineering problems by extracting non-intuitive patterns in complex data. Exploiting non-volatile memory designs for these systems is a natural choice. Few explorations of neuromorphic algorithms were shown in TrueNorth \cite{hsu2014ibm}, NeuroGrid \cite{benjamin2014neurogrid}, and BrainScales \cite{BrainScales2014},
architectures, where the synaptic memory is placed adjacent to the neuron circuits. Overall chip architecture will have multiple neuro-synaptic cores which communicate using different asynchronous and synchronous processing. However, the proposed HTM spatial pooler architecture takes a  different approach, where the hardware design is custom tuned for the spatial pooler algorithm while exploiting the nonvolatile memory.

The aim of this work is to explore a pragmatic architecture for the HTM spatial pooler and evaluate its ability to scale with respect to classification performance, memory limitations, latency, area, and power consumption.
The remainder of this work is structured as follows: In Section \ref{section:spatial_pooler}, a hardware-amenable mathematical formulation of the CLA SP is described. Microarchitecture for a storage processor unit (SPU) implementation, namely Non-volatile HTM (NVHTM), is discussed in Section \ref{section:fhtm}. NVHTM is evaluated for its scalability, based on various design constraints in Section \ref{section:results}. Conclusions and future work may be found in Section \ref{section:conclusion}.

\section{Spatial Pooler Model}
\label{section:spatial_pooler}
HTM SP is a three-phase biomimetic unsupervised clustering algorithm that seeks to create a mapping from an encoded input to an SDR. In order, it consists of overlap, inhibition, and learning phases. These operations are computed by individual nodes, or columns, within a two-dimensional column space (Fig. \ref{fig:htm_region}). Each aspect of the spatial pooling process is progressively discussed, starting with a simple description of overlap, eventually arriving at the full embodiment of SP with a learning rule. Limitations of the model are discussed to identify where NVHTM diverges from CLA.

\subsection{Phase I: Overlap}
HTM columns are nodes in the two-dimensional column space that contain a proximal segment:  synaptic state datastructure.
Overlap describes the degree of correspondence between a given proximal segment and the current feedforward stimulus: the input region. This operation is a dot product between the proximal segment and the input region. Columnar proximal segments may be represented as a vector of real-valued permanences, bounded between zero and unity, inclusive. Once the permanence magnitude is above a threshold, $P_{th}$, a potential synapse is promoted to the connected state. Synaptic connectivity is determined by the condition (\ref{eqn:psyn_thresh}); $\vec{c}_{i}$ represents the proximal segment vector for the $i^{th}$ column 
and $\vec{C}_{i}$ is the binary vector of columnar synapse state: a value of 1 is indicative of a connection.

\begin{equation}
	\label{eqn:psyn_thresh}
	\vec{C}_{i} = \vec{c}_{i}\ge P_{th}
\end{equation}

There is a one-to-one mapping between elements within the column vector and the input region. Given that proximal segments may be represented as length-$K$ vectors, the entire column space may be envisioned as a $K \times N$ matrix, $\mathbf{C}$, 
in which the $i^{th}$ column vector corresponds to the proximal segment of the $i^{th}$ HTM column.
Each of the $N$ columns are capable of being connected to a maximum of $K$ synapses. 

\begin{equation}
	\label{eqn:alpha_i_prime}
	\alpha_{i}' = \vec{C}_{i} \cdot \vec{X}_{t}
\end{equation}

\begin{equation}
	\label{eqn:alpha_prime}
	\vec{\alpha'} = \mathbf{C}^T \cdot \vec{X}_{t}
\end{equation}

Overlap for the $i^{th}$ column, the dot product of its proximal segment state vector and the binary input vector, is modeled by (\ref{eqn:alpha_i_prime}). The entire network may be evaluated in parallel by computing a dot product between the transpose of $\mathbf{C}$ and the current input state vector, $\vec{X}_{t}$, as shown in (\ref{eqn:alpha_prime}); $\alpha_{i}'$ and $\vec{\alpha'}$ are overlap for the $i^{th}$ column and the entire network, respectively. This relationship is extended to account for the boost factor in addition to minimum overlap, yielding

\begin{equation}
	\label{eqn:alpha_i}
	\alpha_{i} = 
	\begin{cases}
		\alpha_{i}'\beta_{i}, \& \alpha_{i}' \ge A_{th} \\
		0, \& otherwise 
	\end{cases}
\end{equation}

where $\alpha_{i}$ is the overlap value for the $i^{th}$ column after boosting and minimum overlap thresholding;
$A_{th}$ is the minimum overlap threshold; and $\beta_{i}$ is the boosting factor. In \cite{Numenta2011}, boost factor is restricted to values greater than or equal to unity.

\subsection{Phase II: Inhibition}
A competitive learning process, referred to as inhibition, is leveraged to select candidate columns to enter the active state--thus contributing to the output SDR. Columns are inhibited by neighbors within their inhibition radius and are pruned if their overlap is below that of the $K^{th}$ largest within their radius, which is defined with respect to column-space coordinates (Fig. \ref{fig:htm_region}). Inhibition is classified into two forms, namely global and local; the latter of which being significantly more complex to compute. 


Inhibition applies a k-select operation and pruning step to remove all columns within a set and an overlap value below that of the selected column. This may be modeled by sorting all columns, $j$, within the inhibition radius of another column, $i$; hereto referred to as $\Lambda_{j}^{i}$. Pruning is modeled in (\ref{eqn:column_activation}) by removing columns below a specified index from the aforementioned set, which now becomes the SP output SDR.

\begin{equation}
	\label{eqn:column_activation}
	\mathcal{A}_{i} = \alpha_{i} \ge \vartheta[i, d_{th}]
\end{equation}

\begin{equation}
	\vartheta[i, d_{th}] = \Lambda_{j}^{i}[min(d_{th}, length(\Lambda_{j}^{i}))]
\end{equation}

where $\mathcal{A}_{i}$ is the binary inhibited columnar activation; $d_{th}$ is the desired local activity; $\vartheta$ is a bounded indexing function--used to index into $\Lambda_{j}^{i}$, subject to $d_{th}$; $\vartheta[i, d_{th}]$ calculates the minimum local activity for the $i^{th}$ column; the $min$ function is used to restrict the range of $\vartheta$ to indexes within $\Lambda_{j}^{i}$.

Global inhibition is a special case of the more generalized local model. Given that the inhibition radius of each column is a known constant, computational complexity may be simplified. The global case is configuration in which the ``\emph{local}" inhibition radii of all columns are large enough to encapsulate the entire column space: a known fixed quantity. Consequently, the equation governing both global and local cases are, in principle, the same. Implementations of SP in the literature utilize global over local inhibition, yielding comparable results \cite{Mnatzaganian2016}. Furthermore, the global case is more practical for hardware implementation, requiring fewer parameters to be tracked, such as inhibition radii, implicitly removing the requirement for updating these parameters dynamically.

\subsection{Phase III: Learning}
\subsubsection{Active Segment Update}
The third, and final, phase of spatial pooling is to train the proximal segment state matrix. SP learning is influenced by network state (due to prior training and observed patterns) in addition to the current feedforward input vector. The current input is dictated by

\begin{equation}
	\label{eqn:prox_seg_ff_update}
	\vec{C}_{i}^{*}[j] = 
	\begin{cases}
		\vec{C}_{i}[j] + \lambda\vec{X}_{t}[j] - P_{dec}, \& \mathcal{A}_{i} = 1 \\
		\vec{C}_{i}[j], \& otherwise 
	\end{cases}
\end{equation}

where $\vec{C}_{i}^{*}[j]$ is the new value for the $j^{th}$ permanence within the $i^{th}$ proximal segment state vector; bounded between 0 and 1, inclusive; $\lambda$ is the sum of $P_{inc}$ and $P_{dec}$--the permanence increment and decrement, respectively. Synaptic state is only updated for active columns and is based upon the binary input region vector. If the bit was active, the permanence is incremented, otherwise it is decremented.

\subsubsection{Duty Cycle Update}
Following active segment update, duty cycle parameters are updated to enforce fairness between columns. Duty cycle is used as a measure of the extent to which a column participates in representing the encoded input region; it is modeled by an active duty cycle, $D_A$, and overlap duty cycle , $D_O$. $D_A$ is a measure of how frequently a column wins in the inhibition phase; $D_O$ is a measure of the average pre-inhibited overlap for the column in question. Active duty cycle is updated such that
\begin{equation}
	D_{A}^{*}[i] = \frac{D_{A}[i](\tau_{D} - 1) + \mathcal{A}_{i}}{\tau_D}
\end{equation}

where $D_{A}^{*}[i]$ is the next value for the active duty cycle, $D_{A}$, of the $i^{th}$ column; 
$\tau_{D}$ is the period over which the duty cycle is evaluated and consequently averaged,
and $\mathcal{A}_{i}$ is the current active state for the column. A similar process is repeated for the overlap duty cycle in (\ref{eqn:overlap_duty_cycle}).

\begin{equation}
\label{eqn:overlap_duty_cycle}
	D_{O}^{*}[i] = \frac{D_{O}[i](\tau_{D} - 1) + \alpha_{i}}{\tau_D}
\end{equation}

$D_{O}^{*}[i]$ is the next value for the overlap duty cycle, $D_{O}$. $\alpha_{i}$ is the new post-inhibition overlap resulting from the current feedforward input. Although active and overlap duty cycles were implemented having the same period. This is not an inherent limitation of CLA. 

\subsubsection{Weak Column Boosting}
Weak column boosting, described by (\ref{eqn:weak_boosting}), seeks to increase the number of synapses that are in the connected state for potentially starved columns. The permanence values of a column, $\vec{C}_{i}$ are increased by 10\% of the permanence threshold \cite{Numenta2011}. Other such magic numbers were used in the white paper, but further work should seek to optimize the HTM to ascertain optimal parameters using parametric optimization techniques such as simulated annealing.

\begin{equation}
	\label{eqn:weak_boosting}
	\vec{C}_{i}^{*}[j] = 
	\begin{cases}
		\vec{C}_{i}[j] + P_{th}/10, & D_{O}[i] < \widetilde{D}_{O}[i], \vec{C}_{i}[j] > 0 \\
		\vec{C}_{i}[j], & otherwise 
	\end{cases}
\end{equation}

where $\widetilde{D}_{O}$ is the minimum overlap duty cycle. If the duty cycle of the column in question is below $\widetilde{D}_{O}$ and the synapse has not already desaturated to a value of 0 (pruned), the permanence values are incremented.

\subsubsection{Boost Factor Update}
The overall goal of boosting is to reduce column starvation, promoting an increase in the number of columns contributing to the process of creating a model of the input data. This aspect of the learning algorithm is modeled as

\begin{equation}
\label{eqn:active_boosting}
	\beta_{i}^{*} = 
	\begin{cases}
		\frac{1-\beta_{max}}{\widetilde{D}_{A}}D_{A}[i] + \beta_{max}, & D_{A}[i] < \widetilde{D}_{A} \\
		\beta_{i}, & otherwise 
	\end{cases}
\end{equation}

where $\beta_{i}^{*}$ is the next boost factor value, $\beta_{i}$, for the $i^{th}$ column; $\widetilde{D}_{A}$ is the minimum activity duty cycle threshold, and $\beta_{max}$ is the maximum boost factor. The system in (\ref{eqn:active_boosting}) is a piece-wise linear function, where $D_{A}[i]$ is the only quantity that varies between iterations. Within the original CLA algorithmic formulation, $\widetilde{D}_{A}$ is also specified as a variable quantity. However, as a hardware simplification, this  parameter is expected to be extracted from a more complete software model before implementing the model on the hardware platform.

\begin{figure}[t]
	\centering
	\includegraphics[width=\linewidth]{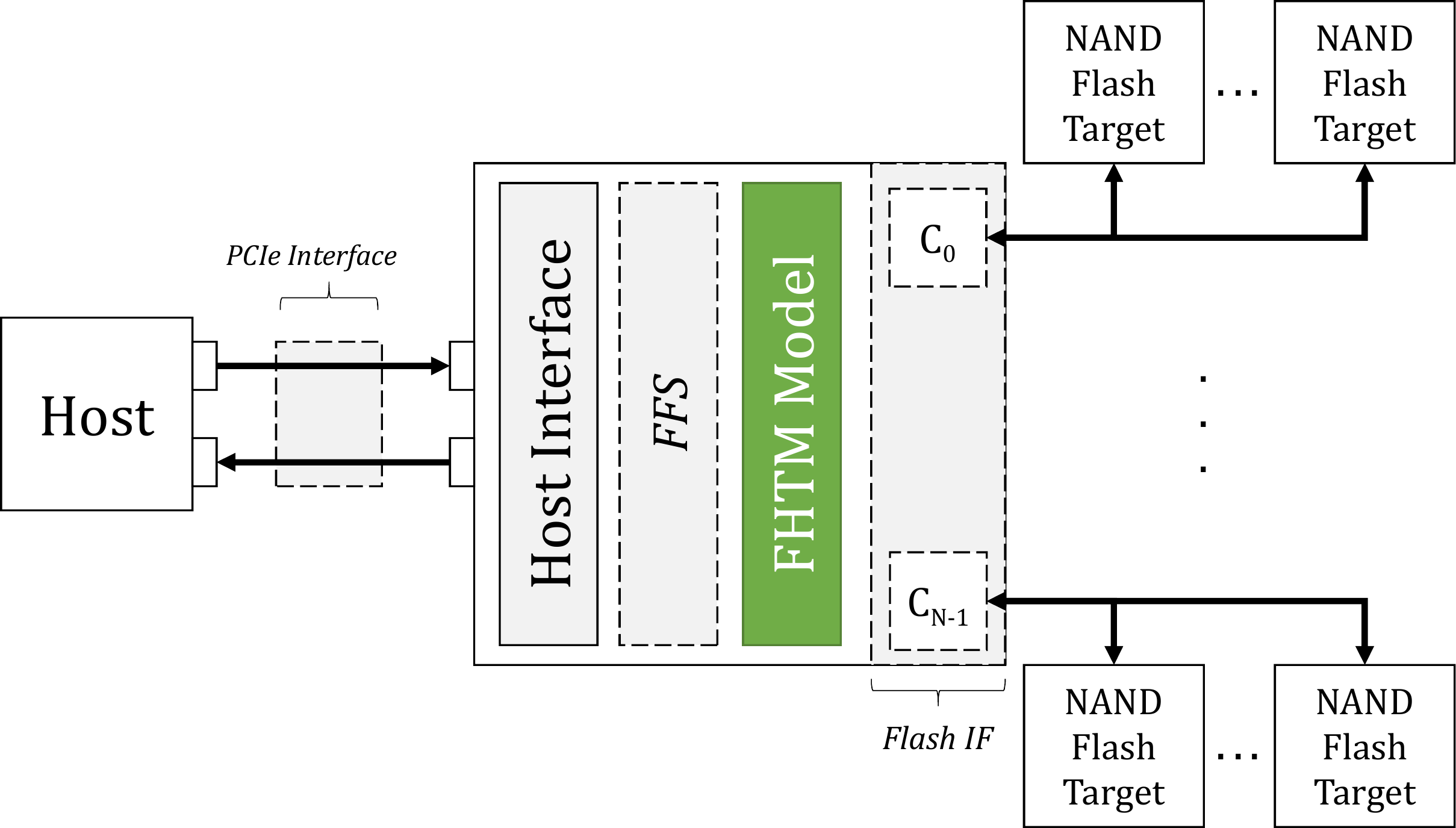}
	\caption[]{High-level NVHTM architecture concept. An SSD with a flash file system for address translation. Host interface to manage interoperability with the PCIe interface standard. NVHTM augments the typical SSD structure by incorporating the accelerator into existing data path buses.}
	\label{fig:htm_highlevel}
\end{figure}

\section{Proposed NVHTM Architecture}
\label{section:fhtm}
The NVHTM architecture, implemented in VHDL and depicted in Fig. \ref{fig:htm_highlevel}, was used to extract area, power, and latency estimates. Operations defined by the mathematical formulation were modeled in hardware as a parallel, pipelined SPU, targeting solid-state drive (SSD) storage devices. Our proposal is that HTM SP logic may be added to the read/write data path, to enable improved parallelism and greater potential for model scalability. Pages of data, containing proximal segment state, are stored within NAND flash target devices through flash channel interfaces. An in-system flash file system (FFS) manages scheduling read and write operations; the host interface exists to translate data from a host CPU to a series of flash commands. Within the hardware model only the NVHTM data path is presented.

\begin{figure}[t]
	\centering
	\includegraphics[width=\linewidth]{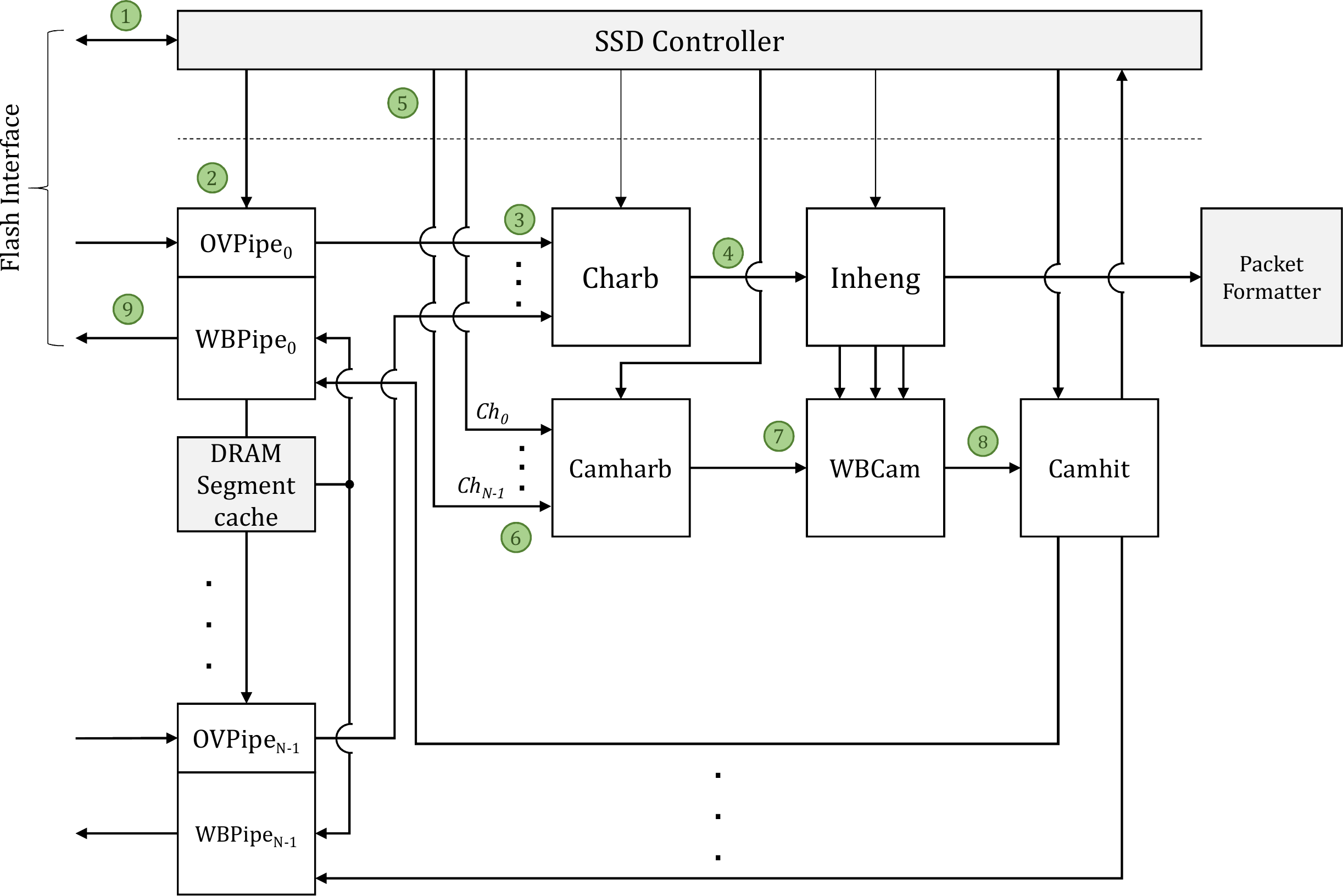}
	\caption[Microarchitecture for NVHTM SP]{Microarchitecture for NVHTM SP. Number labels have been applied to indicate the flow of data through SP with learning enabled. Grayed out components indicate that these were not implemented in the NVHTM design but are described to establish context.}
	\label{fig:fhtm_spu}
\end{figure}

\begin{figure}[t]
	\centering
	\includegraphics[width=\linewidth]
	{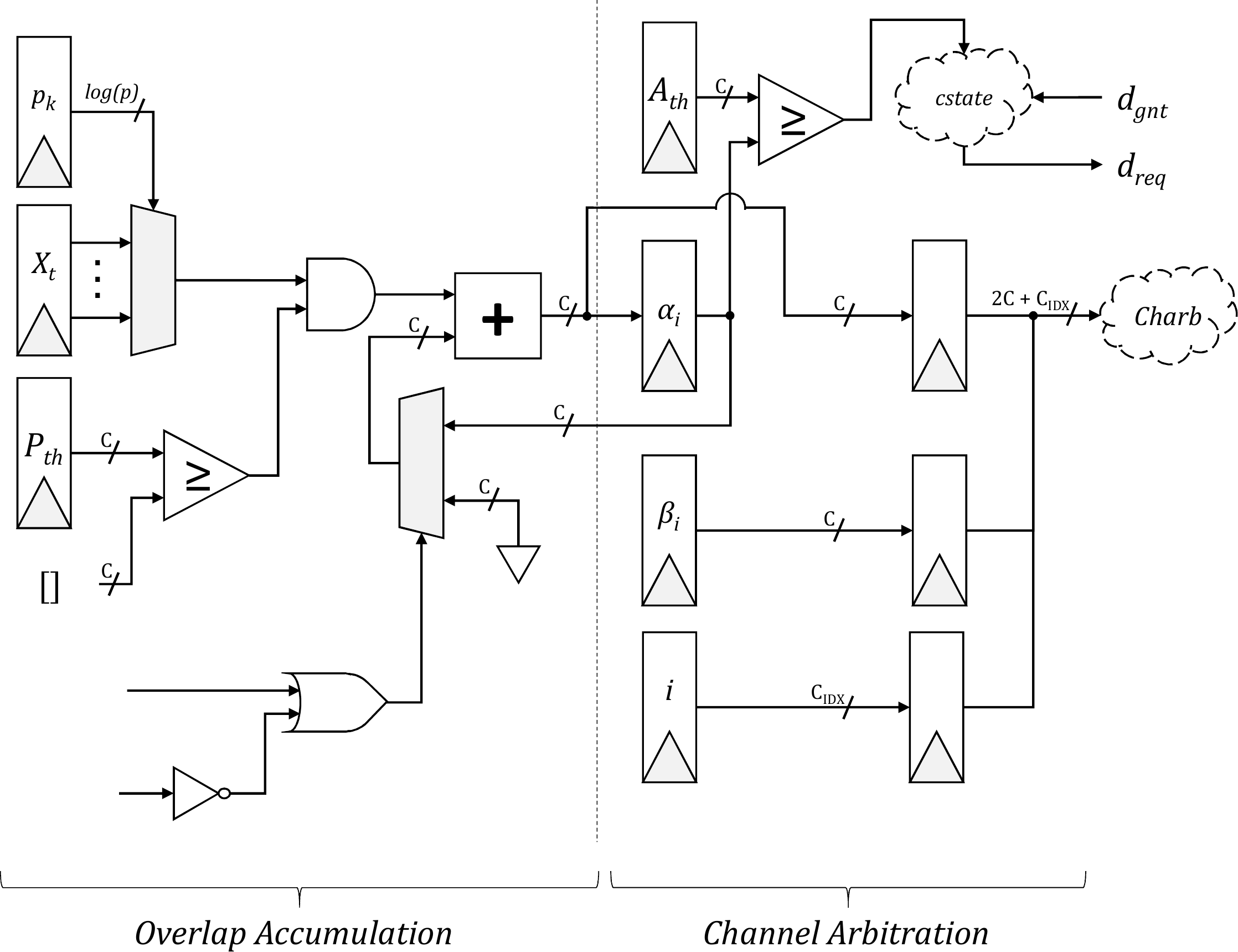}
	\caption[Overlap pipeline for NVHTM architecture]{Overlap pipeline for NVHTM architecture. Configuration data is stored in each of the registers: $P_{th}$ is the permanence threshold; $C_{i}[k]$ is the permanence for the $k^{th}$ proximal synapse of the $i^{th}$ column; $A_{th}$ is the overlap threshold; $\beta_i$ is the boost constant; $k$ is a pointer used to index into the input vector; $p$ is the length of $X_t$; $p_k$ is the pointer used to index $X_t$; $C$ is the channel width; $C_{IDX}$ is the number of bits used to encode the column index; $psv_0$ is a pipeline stage enable signal; $d_{gnt}$ and $d_{req}$ are the arbitration signals; and $d_{val}$ marks the first valid data on the pipeline. Output of the overlap pipeline is written to a DRAM table and to the inhibition engine. The in-DRAM overlap table is used within the learning phase to update the duty cycle and boost parameters.}
	\label{fig:vmlx_overlap}
\end{figure}

Spatial pooling in NVHTM is modeled by augmenting the data path with an overlap pipeline \emph{OVPipe}, inhibition engine \emph{Inheng}, and learning subsystem (\emph{WBPipe} and \emph{WBCntl}). NVHTM auxiliary functions include the channel arbiter (\emph{Charb}), content-addressable memory arbiter (\emph{Camharb}), write-back content addressable memory (\emph{WBCam}), and CAM hit detector (\emph{Camhit}); \emph{WBCntl} is comprised of \emph{Camharb}, \emph{WBCam}, and \emph{Camhit}. NVHTM microarchitecture, depicted in Fig. \ref{fig:fhtm_spu}, operates in the following manner:
\begin{enumerate}
	\item The SSD controller initiates operation by issuing a read command to the flash channel.
	\item The overlap pipeline (\emph{OVPipe}) is enabled along with the other functional pipes in the read path. Provides a pointer to the input region vector and configuration data broadcast by the host. Data then arrives from the flash interface one beat at a time.
	\item Overlap values are selected by the channel arbiter (\emph{Charb}) for boosting and inhibition.
	\item  After the single columnar overlap and segment state information traverse the pipeline to this point, the inhibition engine (\emph{Inheng}) uses an insertion sort operation to determine the subset of columns that will enter the active state. 
	\item In the case when learning is enabled, the controller issues column indexes to the content addressable memory arbiter (\emph{Camharb}).
	\item \emph{Camharb} then arbitrates between channels to select a column index to be sent downstream.
	\item The write-back content addressable memory (\emph{WBCam}) is used to determine which columns are candidates for learning by comparing the issued column index against valid indexes acquired in the inhibition phase.
	\item Hits in the CAM are detected (\emph{Camhit}) and used to control the write-back ALU (\emph{WBPipe}). Timeout signals are forwarded back to the SSD controller to indicate a CAM miss.
	\item \emph{WBPipe} uses input from \emph{Camhit} to control how the proximal segment, which is cached in the segment cache, will be written back to memory through the flash interface. State information within the inhibition engine is used to select operands for the ALU within \emph{WBPipe}, which updates each segment one synapse at a time.
\end{enumerate}

\begin{figure}[t]
	\centering
	\includegraphics[width=\linewidth]
	{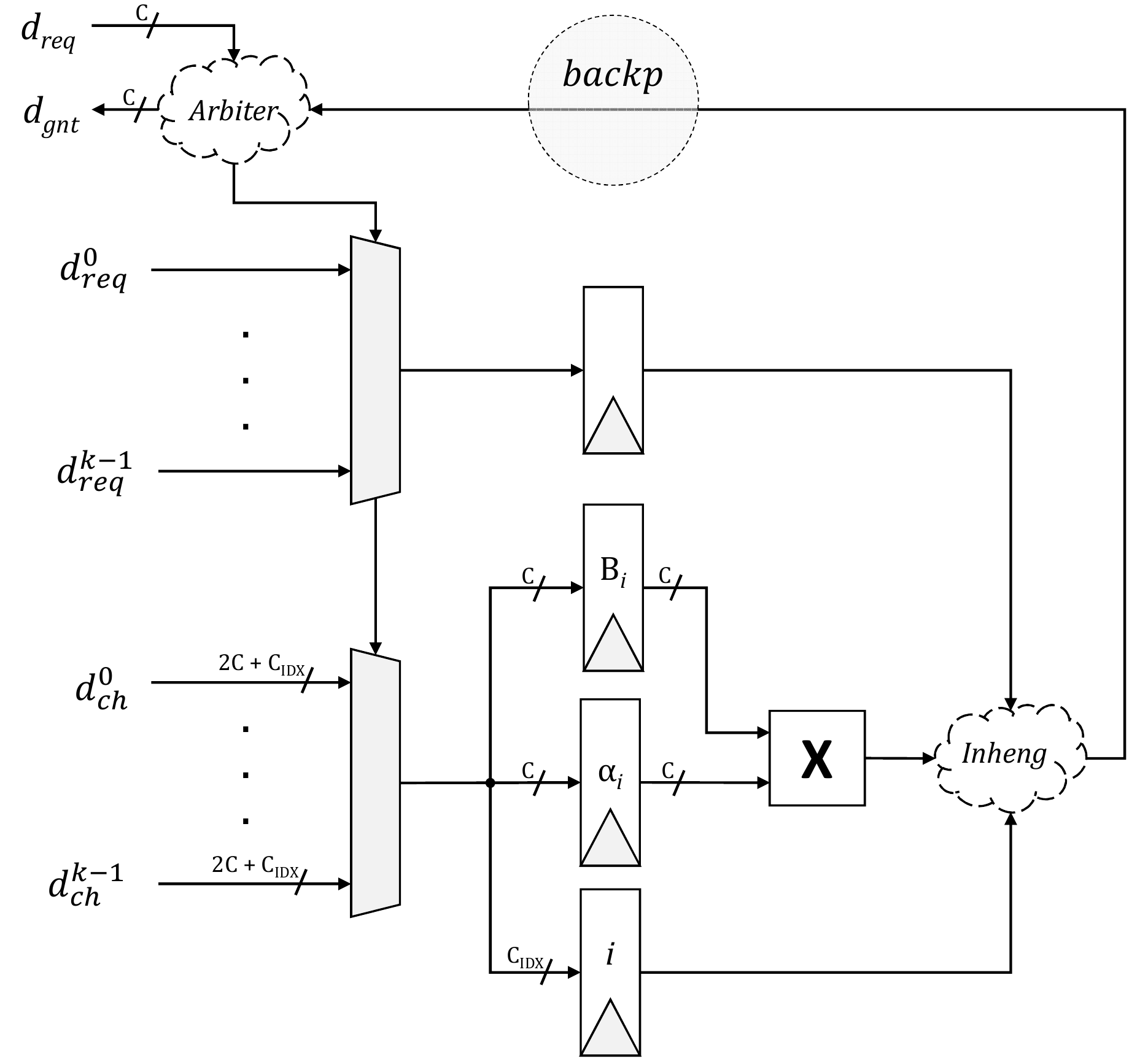}
	\caption[The channel arbiter manages transmitting overlap values from multiple overlap engines to the downstream inhibition engine]{The channel arbiter (\emph{Charb}) manages transmitting overlap values from multiple overlap engines to the downstream inhibition engine. Furthermore, the overlap values accumulated in \emph{OVPipe} are boosted at this phase prior to being distributed to the inhibition engine. The selection of proximal segment data results in the associated channel receiving a data grant acknowledgment, $d_{gnt}$.}
	\label{fig:fhtm_charb}
\end{figure}

After the NVHTM architecture has been configured by the host processor, the SPU waits for an input stimulus, $X_t$, to be broadcast to the SSD. The host processor transmits a packet containing the input vector. Upon reception of the input vector, the FFS generates commands that signal the NAND interface to supply pages of data corresponding to stored proximal segments. Each beat of data on the interface corresponds to synaptic permanences and other SP column states. The codewords of data sourced from the NAND, after low-density parity-check (LDPC decode), are supplied to \emph{OVPipe} one beat of data at a time.

The overlap pipeline, depicted in Fig. \ref{fig:vmlx_overlap}, is an accumulator that operates synchronous to the flash interface clock. Once the first beat of data arrives at the overlap engine from the LDPC, all subsequent beats of data will become available in the immediately following cycles. Furthermore, because data being read has a predictable structure, a drastically simplified overlap engine was designed. Additional simplification is obtained by exploiting the simplicity of mathematical constructs used by the algorithm.

Latency on the order of tens of cycles are added to the data path due to the vector processor. To put this into perspective, the number of cycles required for the read and write operations are on the order of hundreds to thousands of cycles. Taking into account the interface limitations, additional pipeline complexity is likely to have a negligible impact on performance. As design complexity scales, resulting in larger page sizes, more channels, and higher data rates, this simple model will continue to be feasible and effective. This is because the pipeline has few opportunities for significant critical path limitations: (1) the data interface is only 8-bits to 16-bits wide in most flash implementations, leading to lower computation delays than common 32-bit and 64-bit hardware; (2) there are a small number of inputs to each logical component; and (3) operating at the memory interface frequency provides a large upper bound on allotted cycle time.

\begin{figure}[t]
	\centering
	\includegraphics[width=\linewidth]
	{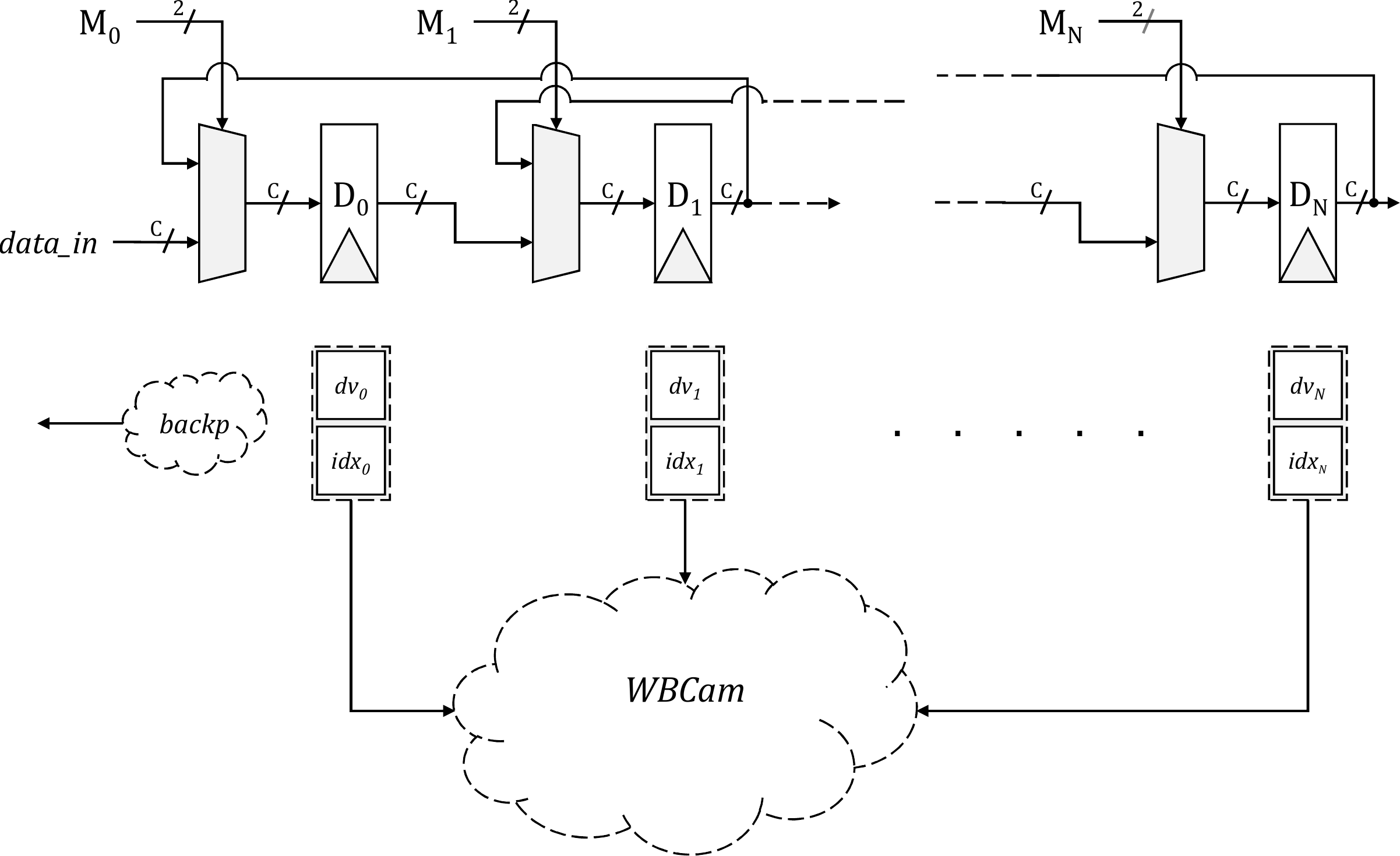}
	\caption[The inhibition engine for the NVHTM architecture]{The inhibition engine for the NVHTM architecture. Control signals are generated by the swap state registers in conjunction with data valid bits and the results of comparison operations. The inhibition engine state is cleared upon the arrival of a new input vector, $X_t$. Note that the column index data path is abbreviated for clarity--it follows a similar path to that of the overlap overlap swap logic, $D_i$.}
	\label{fig:vmlx_inhibition_eng}
\end{figure}

As a new $X_t$ input vector arrives from the host, a pointer, $j$, is reset to reference the first bit of the vector. The multiplexer is utilized to select the respective bit within the vector. This first portion of the overlap engine (separated by a dashed line) computes the pre-boosted overlap for the column using an accumulator. Input data arriving from the host is one-to-one matched with the number of potential candidate synapses. Each synaptic value arriving from the flash interface is thresholded using a comparator. An AND gate determines that an active-connected synapse was found. In the event that a proximal synapse is both active and connected, the overlap value is incremented. The internal state of the accumulator is reset whenever a new input vector arrives. This complete process takes an amount of time proportional to the page size divided by the channel width--it spans the time required for the read operation plus the overlap pipeline latency.

Following data accumulation in \emph{OVPipe}, the channel arbiter, shown in Fig. \ref{fig:fhtm_charb}, is notified that valid data is ready for boosting and inhibition. Only overlap values that have overcome the minimum-overlap constraint issue a valid data request to the downstream logic. A full handshake is computed using a request-grant interface. The purpose of this arbitration is to ensure that only one overlap value is sent downstream and that the other channels are stalled until the inhibition engine is able to service each request in order.

\begin{figure}[t]
	\centering
	\includegraphics[width=\linewidth]
	{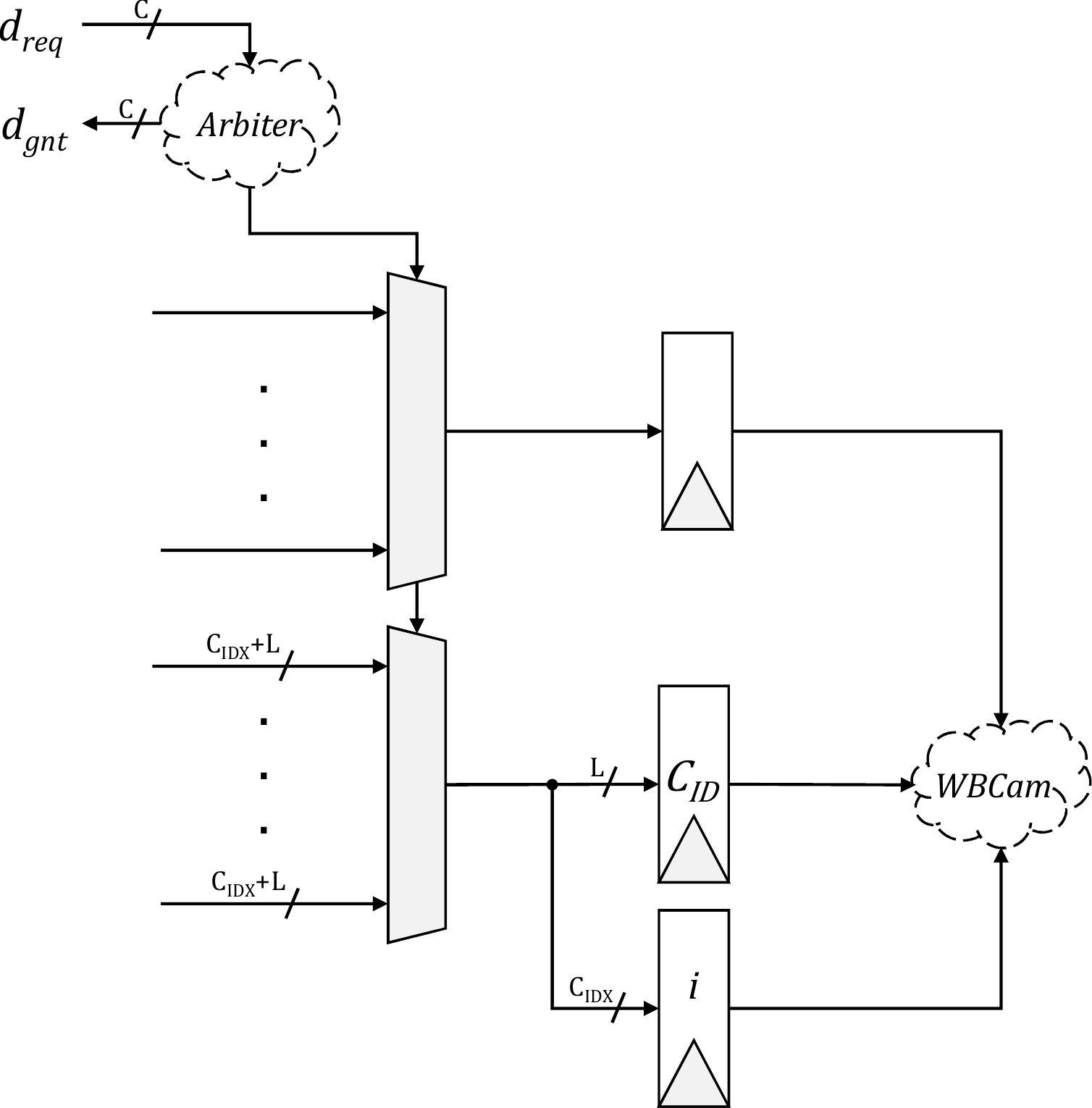}
	\caption[The content-addressable memory channel arbiter]{The content-addressable memory channel arbiter (\emph{Camharb}) operates similarly to the \emph{Charb}, but does not implement back-pressure because the downstream logic cannot be overrun as with the inhibition engine.}
	\label{fig:sp_camharb}
\end{figure}

Inhibition for SP is modeled as an insertion sort in which values within the sorted list below a predefined index are discarded. To accomplish this, a pipeline capable of sorting elements in linear time was created. \emph{Inheng}, depicted in Fig. \ref{fig:vmlx_inhibition_eng}, is represented as a multi-directional shifting queue. Overlap values generated from the activation due to the feedforward input are transferred to the inhibition engine after being selected by the channel arbiter. As the overlap pipeline operates on the page data, the inhibition engine attempts to place, and subsequently sort it within the queue structure. Considering the fact that overlap requires hundreds of cycles to compute, the inhibition engine is allotted a large window of time to complete each sort operation.  This lead to the design decision to make the sort engine linear, minimizing hardware complexity, whilst still adhering to timing constraints.

This design assumes that the large number of cycles required to compute the overlap eliminates any possibility of overrunning the channel arbiter and in-turn, the inhibition engine. \emph{Inheng} applies back-pressure to \emph{Charb}, stalling it until the prior sort operation has completed. Back pressure acts in a flexible way such that the complete pipe need not be stalled by downstream blockage. Pressure is funneled back upstream stage-by-stage, allowing gaps in the pipe to be filled where possible. For example, if \emph{Inheng} is applying back-pressure and the output of arbitration is invalid, valid data may still move forward in the pipe.

Data being shifted around in the inhibition queue is $C+C_{IDX}$ bits wide; where $C$ is the flash interface width and $C_{IDX}$ is the number of bits required to encode the column index. The linear sort implemented by the inhibition engine is composed of multiple complementary swap operations governed by sequential control logic. Each swap operation is controlled using a two-bit signal, $M_i$, generated by a swap state register in conjunction with additional combinational control logic. The most significant bit of the swap indicates direction, while the other bit is utilized as a functional enable. Data elements within the queue are swapped using a multiplexer.

Valid bit, column index, and overlap values are moved through the pipeline, controlled by the comparison of overlap data. As new data is loaded, swap requests propagate through the queue until a valid swap operation is no longer possible. The swap logic checks for validity between the two elements and compares their magnitude, taking into account queue fullness. \emph{Inheng} is deactivated after completing the swap operation as it waits for new overlap data to become available.

\begin{figure}[t]
	\centering
	\includegraphics[width=\linewidth]
	{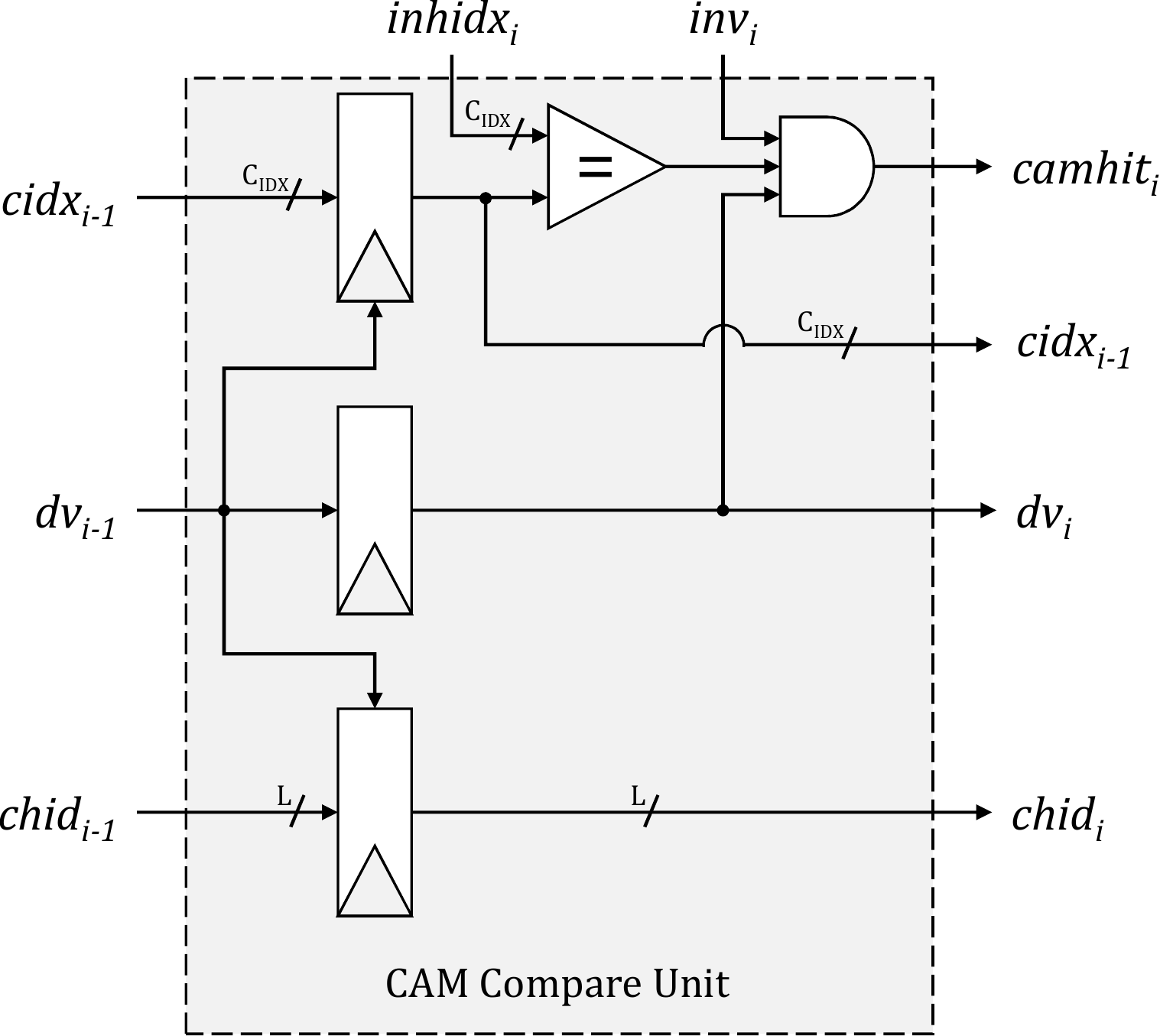}
	\caption[The fundamental unit that is chained together to form the \emph{WBCam}]{The fundamental unit that is chained together to form the \emph{WBCam}. Column index ($cidx_{i-1}$), data valid ($dv_{i-1}$), and channel id ($chid_{i-1}$) from prior stages in the \emph{WBCam} logic are fed into the current stage of the pipeline. Column indexes found in the inhibition queue ($inhidx_{i}$) are applied to the pipeline and are qualified by valid signals ($inv_{i}$) to support partially filled inhibition queues.}
	\label{fig:sp_wbcam}
\end{figure}

The control signals are specified to allow all registers to shift in the same direction when new data is loaded, and to subsequently sort that data. After all HTM columns have been processed by \emph{Inheng}, the column indexes corresponding to the overlap values stored within the queue are used to generate an SP output packet due for transmission to the host processor. Column indexes generated by the inhibition phase are also used in the learning phase to influence proximal segment update.

Read data for SP flows from \emph{OVPipe}, to the \emph{Charb}, finally ending in \emph{Inheng}, which supplies output to a packet formatter that emits through the host interface. Learning follows a slightly different data flow and must immediately follow the completion of inhibition for all columns. Indexes for all columns in \emph{Inheng} are fed into the \emph{WBCam} for use in the learning phase. Proximal segment update is initiated by the SSD controller, which must schedule reads for all columns in the network and also issue corresponding notifications to the content-addressable memory arbiter (\emph{Camharb}), shown in Fig. \ref{fig:sp_camharb}. The SSD controller is notified of hits and misses in the CAM by \emph{Camhit}.

\emph{WBCam} is comprised of chained compare units, depicted in Fig. \ref{fig:sp_wbcam}. For each stage other than the first, if a hit is detected, the data is invalidated in the next stage in the pipeline to avoid another subsequent comparison. This feature is requisite to avoid invalid detection of timeouts in the pipeline. For all compare components other than the first ($i > 0$), the data valid output is defined by (\ref{eqn:sp_dvalid}).

\begin{equation}
	\label{eqn:sp_dvalid}
	dv_{i} = dv_{i-1} \land \overline{inhv_{i-1} \land (cidx_{i-1} = inhidx_{i-1})}
\end{equation}

Proximal segment data is read from the memory, and subsequently cached in the lower latency DRAM. Upon completion of the read operation, the data is written from the segment cache in DRAM back to the SSD NAND array through \emph{WBPipe}. Operands for the write-back operation are selected based upon whether a hit was detected in \emph{WBCam}, and upon the overlap duty cycle. Proximal segment state information is updated in the \emph{WBPipe} as well. In this scheme, the memory can queue up several reads to occur in order, followed by segment updates.

\begin{figure}[t]
	\centering
	\includegraphics[width=\linewidth]
	{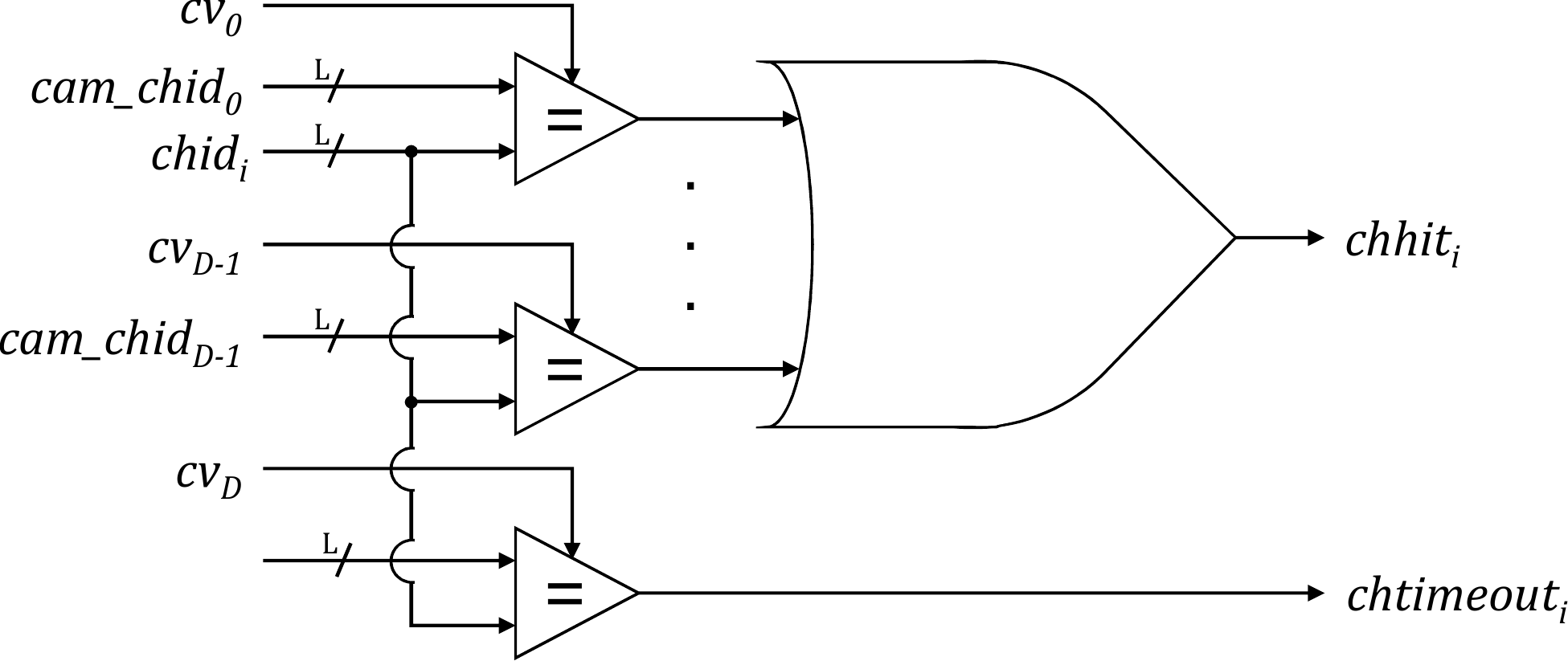}
	\caption[The \emph{Camhit} component is comprised of comparators used to determine whether the ID for the $i^{th}$ channel has been matched in the CAM]{The \emph{Camhit} component is comprised of comparators used to determine whether the ID for the $i^{th}$ channel, $chid_{i}$, has been matched in the CAM. It returns a notification to the SSD controller indicating the event. Valid signals, $cv_{0..D}$, are used to qualify each comparison. A timeout signal, $chtimeout_{i}$, may also be asserted indicating that a compare operation timed out, resulting in a CAM miss.}
	\label{fig:sp_camhit}
\end{figure}

Hits in the CAM are associated with their originating channel via the \emph{Camhit} component, which utilizes a series of equality comparators to aggregate hits corresponding to a channel of interest. This scheme also follows the assumption that each channel will only issue a single column index into \emph{WBCam} at a time to remove the possibility for ambiguity with respect to detected hits. An additional comparison is made with the final phase of the \emph{WBCam} and this index is assigned to the timeout operation logic. Column indexes are indicated as having timed out if they have been compared against all data in the inhibition engine without having matched any other data. A hit detector unit for a single channel is depicted in Fig. \ref{fig:sp_camhit}. In an SSD with $N$ channels, this logic is repeated $N$ times.

Using the results of CAM hits and misses, the SSD controller modifies the input operands to the \emph{WBPipe} ALU. A hit in the CAM indicates that the column was active and therefore a valid candidate for learning. All segments are still re-read and written back to memory in the event that an update is pending for the column. Segments may be updated due to being in the active state following feedforward input or as a result duty cycle updates. In the case of an active duty cycle, the whole segment need not be updated, only the active duty cycle and boost factor. Columnar activity and overlap duty cycle result in an update being made to the entire set of synapses on the segment. Duty cycle parameters cached in the DRAM as proximal segments are first read from memory by the \emph{OVPipe} to control the scheduling of segment updates.

\begin{figure}[t]
	\centering
	\includegraphics[width=\linewidth]
	{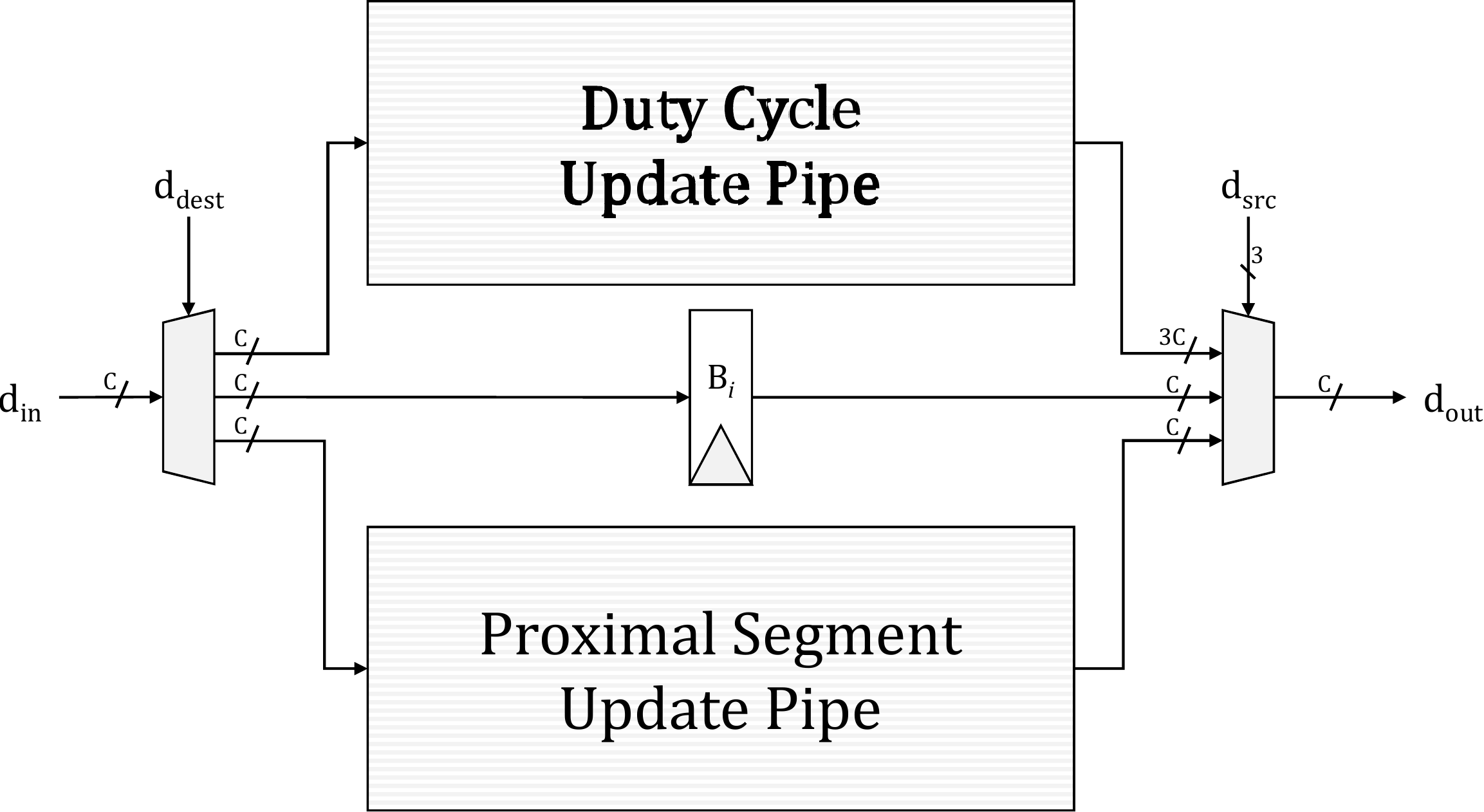}
	\caption[High-level view for the write-back pipeline, \emph{WBPipe}]{High-level view for the write-back pipeline, \emph{WBPipe}. This structure enables data from the flash interface to be steered toward either the duty cycle or proximal segment update pipe. Furthermore, an additional staging register is added to forward the original boost factor during its update phase to allow for the new boost value to be rejected in the event that an active duty cycle is above threshold. Sequencing is controlled by the SSD controller, allowing data to be sourced from either of the aforementioned pipes.}
	\label{fig:sp_wbpipe}
\end{figure}

State information regarding the proximal segment is updated by the write-back data pipeline shown in Fig. \ref{fig:sp_wbpipe}. This pipeline is capable of updating segment duty cycle and boost parameters in addition to the synaptic permanences on the segment itself. Flash interface data is redirected to the pipeline associated with control information provided by the SSD controller. With this topology, data can be streamed in and out of the update pipe at the flash interface speed to reconstruct the original proximal segment page, given the correct control sequence. Output from \emph{WBPipe} may be sourced from the duty cycle pipeline, the old boost factor value register, or the proximal segment update data path.

Configuration inputs, labeled $y_{1..9}$, each correspond to various pre-calculated constants necessary to compute updates in the learning phase. Each constant is defined in (\ref{eqn:cfg_y}). The segment update formulas were modified to be more amenable to the NVHTM design concept, meaning that variable parameters were separated out and divisions were represented as pre-calculated constants instead of in the hardware. This modification made it possible to eliminate the need for dividers and also reduced the number of multipliers to one per write-back pipeline. Although, this could be improved further by using an out-of-order execution core, capable of issuing commands from a fixed number of agents to a fixed set of resources. Wavefront allocators have been employed in the literature to provide effective resource management in hardware architectures.

\begin{subequations}
	\label{eqn:cfg_y}
	\begin{equation}
		y_{1} = (\tau_{D}-1)/ \tau_{D}
	\end{equation}    
	\begin{equation}
		y_{2} = 1/\tau_{D}
	\end{equation}    
	\begin{equation}
		y_{3} = (1/\widetilde{D}_{A})(1-\beta_{max})
	\end{equation}    
	\begin{equation}
		y_{4} = \beta_{max}
	\end{equation}    
	\begin{equation}
		y_{5} = P_{th}/10
	\end{equation}    
	\begin{equation}
		y_{6} = P_{inc}
	\end{equation}    
	\begin{equation}
		y_{7} = P_{dec}
	\end{equation}    
	\begin{equation}
		y_{8} = P_{inc} + P_{th}/10
	\end{equation}    
	\begin{equation}
		y_{9} = P_{dec} + P_{th}/10
	\end{equation}
\end{subequations}

\section{Results}
\label{section:results}

\subsection{Classification}
\label{section:classification}

\begin{figure*}[t]
	\centering
	\includegraphics[width=\linewidth]
	{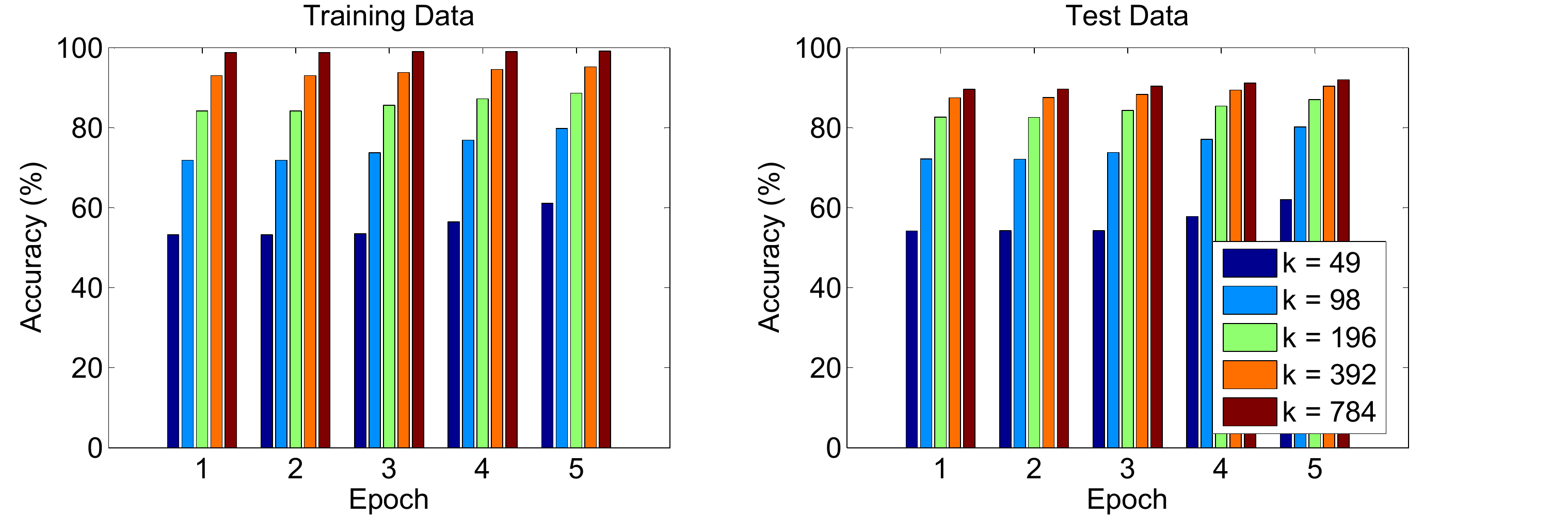}
	\caption[Classification results for the SP]{Classification results for the SP math model, with varying network sizes, $K$. Network performance attained near optimality after the first epoch, with nominal improvement through subsequent iterations.}
	\label{fig:sp_classificaton_results}
\end{figure*}

Results for SP were obtained by simulating the mathematical model using a single thread process on a fixed platform. Execution time required to simulate five epochs across five unique, randomly initialized networks grew from approximately two hours ($K = 49$) to about a week ($K = 784$); this time included training, test, and data collection times for the spatial pooling algorithm. The Matlab model was not optimized for speed, instead, is a demonstration of the functional fidelity of the model.

Classification performance, shown in Fig. \ref{fig:sp_classificaton_results}, was found to reach near optimality after the first epoch, following the law of diminishing returns for subsequent epochs. Accuracies of 99.17\% for training and 91.98\% testing were observed for $K = 784$. The impact of training for additional epochs appeared to play a more significant role in networks with fewer columns. This suggests that an SP-SVM (support vector machine) configuration may be trained with a single SP training epoch for larger network sizes, while still achieving comparable performance. There also seems to be an upper limit on the benefits of network scaling; the improvement from 392 to 784 columns was significantly smaller than that observed between 49 and 98 columns.




\subsection{Network Model Scalability}
NVHTM has the potential to represent large scale networks constrained by the memory size and network configuration. Taking model configurations into consideration, a function was derived to ascertain the potential network size. Total memory and page sizes were used as key parameters. The number of distal segments was considered to demonstrate how this model could be expanded for TM computation support. Maximum network size is limited to the number of pages, because each page represents a single proximal or distal segment. The number of pages within the memory can be simply defined as

\begin{figure}[t]
	\centering 
	\includegraphics[width=\linewidth]{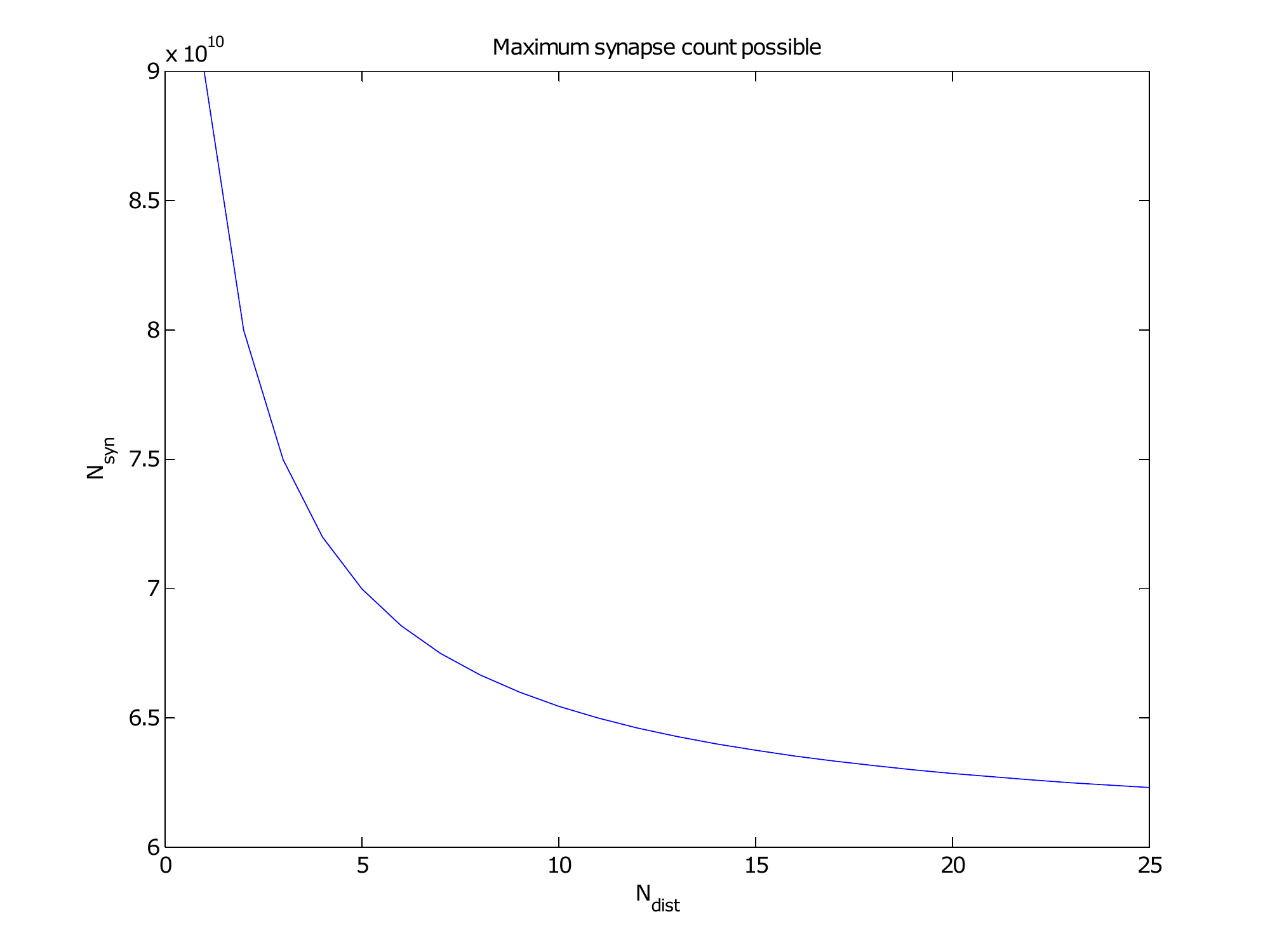}
	\caption[Maximum number of synapses possible for HTM network, as a function of segment size]{Maximum number of synapses possible for HTM network, as a function of segment size, assuming a fixed memory size ($Mem_{sz} = 240GB$), and fixed page size ($P_{sz} = 32kB$); channel width assumed to be 16-bits; column index width, $C_{IDX}$ is assumed to be 24 bits--the number of cells that may be accessed is on the order of millions.}
	\label{fig:max_segments_possible}
\end{figure}

\begin{equation}
	N_{pages} = \frac{Mem_{sz}}{Page_{sz}}
\end{equation}

where $Mem_{sz}$ is the size of user memory space in bytes and $Page_{sz}$ is the size of pages in bytes. Each column has a set of distal segments associated with it, all of which are connected to cells within the respective column. Furthermore, columns are assumed to only have one proximal segment assigned to them. This is used to determine the number of proximal segments as a function of memory size using the relationship

\begin{equation}
	N_{prox} = \frac{N_{pages}}{1 + N_{dist}}
\end{equation}

where $N_{prox}$ is the number of proximal segments; and $N_{dist}$ are the number of distal segments per column: equivalent to the cells per column multiplied by the number of distal segments per cell. Each column is assumed to have a single proximal segment.

Using these equations, the quantity of synapses may also be defined by noting that each proximal segment page has a fixed number of synapses three words less than the page size:the first three words of data contain the overlap and active duty cycles, in addition to the boost factor. Distal segments, assumed to be represented in a sparse format, may only represent a number of synapses less than half the page size (in words), because synapse values are stored in index-value pairs \cite{Zyarah2015}. This is defined using the following system

\begin{equation}
	\label{eqn:nsyn_tlm}
	N_{syn} = N_{prox}\frac{P_{sz}-3C}{C} + N_{prox}N_{dist}\frac{P_{sz} - C}{C + C_{idx}}
\end{equation}

where $N_{syn}$ is the number of synapses within the model; $P_{sz}$ is page size in bytes (flash interface width); $C$ is the number of bytes required to represent a synaptic permanence, assumed to be the flash data interface width (8bits/16bits); $C_{idx}$ is the number of bytes required to represent the cell index for the sparse distal segments. A sample plot of (\ref{eqn:nsyn_tlm}) is shown in Fig. \ref{fig:max_segments_possible}.

\subsection{Power, Area \& Latency}
Hierarchical synthesis was explored for \emph{WBPipe} and \emph{WBCntl} to make synthesizing these larger designs more feasible. Furthermore, instantiated components were compacted to reduce area and an iterative workflow was used to ensure that DRC and LVS requirements were met. More aggressive routing strategies such as wrong-direction routing, ripping and congestion analysis, were required to ensure that all of the overflows between nets would be properly routed. Furthermore, some primary I/O pins had to be manually moved to ensure routability. \emph{WBCntl}: composed of \emph{Camharb}, \emph{WBCam}, and \emph{Camhit}: was created to explore block-based hierarchical synthesis. The power results for each of these components were obtained by extracting the parasitic netlist of \emph{WBCntl}.

\begin{figure}[t]
	\centering 
	\includegraphics[width=\linewidth]{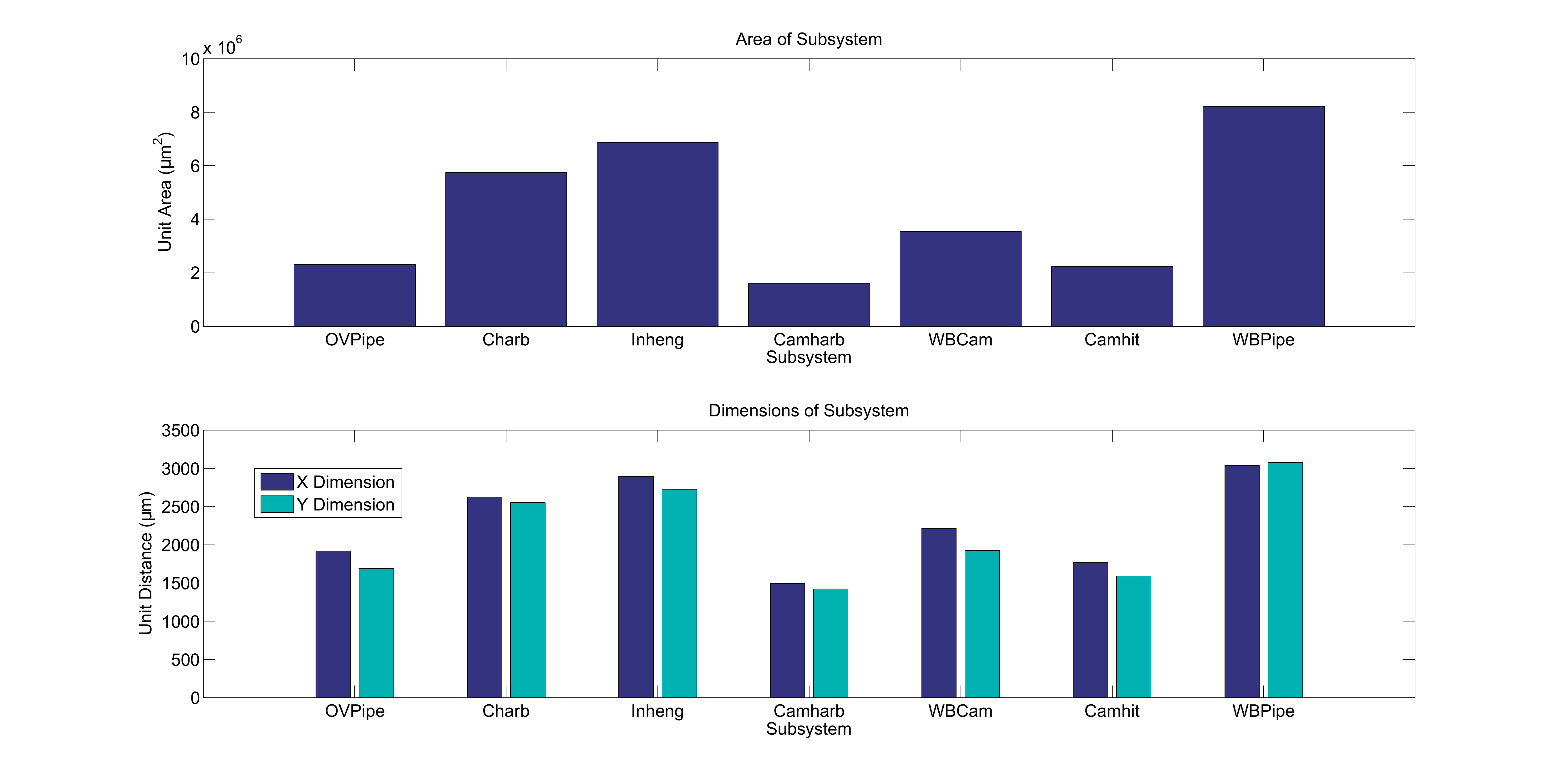}
	\caption[Area and dimensions for each of the synthezied RTL components]{Area (top) and dimensions(bottom) for each of the synthesized RTL components. The read and write pipe logic are expected to consume the greatest amount of area, because they are duplicated per channel, resulting in an 8x-20x increase in area for these components.}
	\label{fig:synth_area}
\end{figure}

Area footprints were generated by Mentor Graphics Pyxis IC design suite as summarized in Fig. \ref{fig:synth_area}. The dimensions of each component are provided, with a total footprint of $30.538mm^2$. However, the estimated area required for an actual implementation must take into account the number of times that each component is expected to be instanced in an actual SSD NVHTM. Estimated area is defined as

\begin{equation}
	A_{NVHTM} = N_{ch}(x_{0} + x_{1}) + x_{2} + x_{3} + x_{4} + x_{5} + x_{6}
\end{equation} 

where $N_{ch}$ is the number of channels, $x_0$ and $x_1$ are the areas for \emph{OVPipe} and \emph{WBPipe}, respectively; the other $x_i$ terms correspond to each of the other components. Under this assumption, the area becomes $104.26mm^2$.

\begin{figure}[t]
	\centering 
	\includegraphics[width=\linewidth]{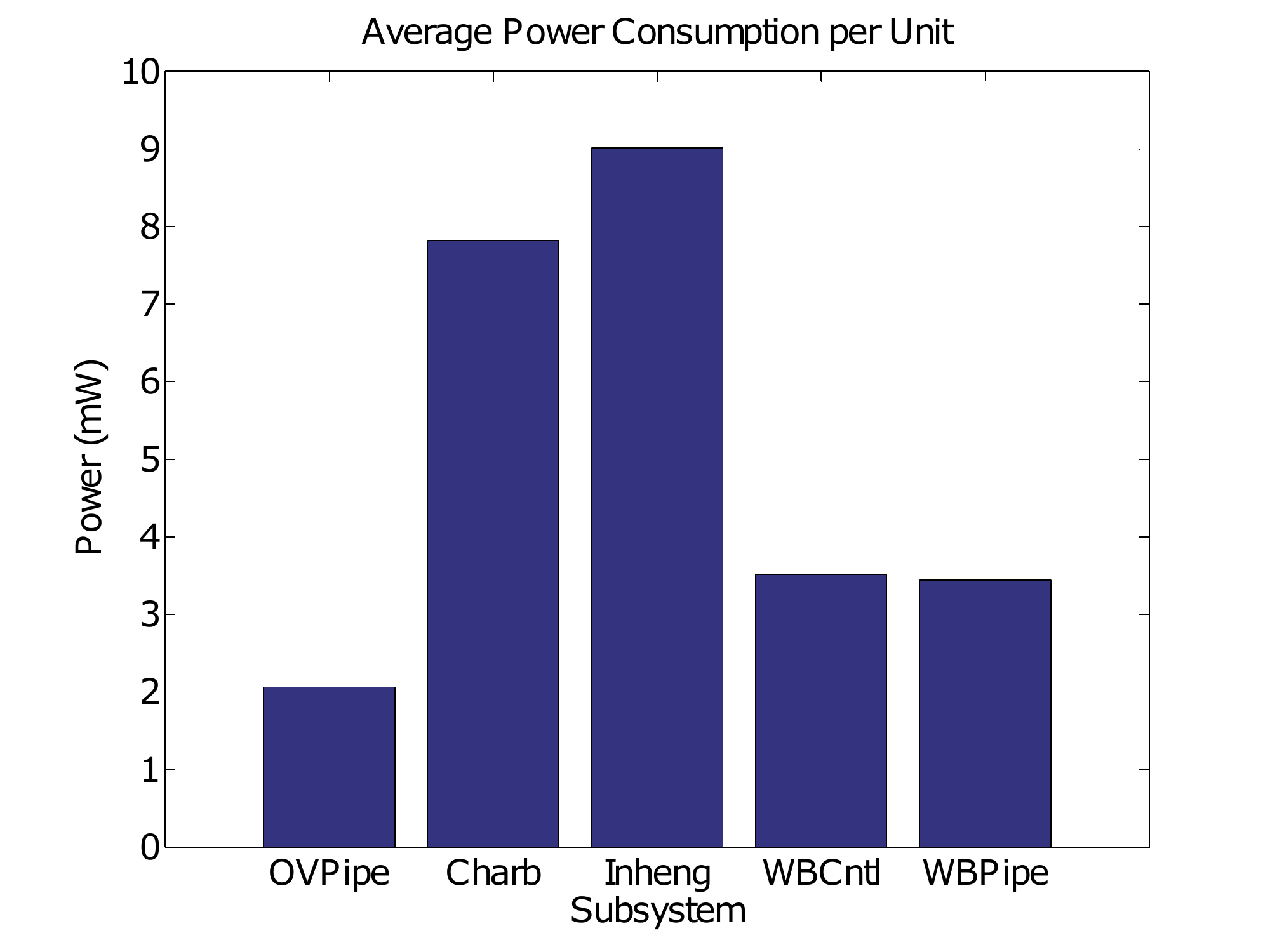}
	\caption[Average power consumed for each of the synthesized RTL components]{Average power consumed for each of the synthesized RTL components.}
	\label{fig:synth_power}
\end{figure}

A parasitic extraction netlist was derived from the layout and evaluated using the Eldo analog simulator. Hardware created at this phase was found to consume power on the order of a few milliwatts, as depicted in Fig. \ref{fig:synth_power}. Components consuming the most power did so due to having a larger number of memory elements switching state at each clock cycle. Average power consumed per component was $5.171mW$--the estimate for an 8-channel SP model is $64.394mW$.

\section{Conclusion \& Future Work}
\label{section:conclusion}
As our understanding of the underlying mechanisms that drive intelligence continues to expand, applications that leverage this body of knowledge will increase. It is likely that machine intelligence, applied to enterprise-scale big data analysis of unstructured data, will serve as catalyst for future innovations. Advances in algorithms will, in turn, lead to the creation of new architectures. Storage processors present a promising opportunity for product differentiation in this space. Companies may provide robust virtual platforms capable of handling pattern recognition workloads at scale with low cost.

Through this exploration, a conceptual design and analysis have been presented for an NVHTM spatial pooler. The impact of augmenting a storage unit with processing capabilities degrades the upfront latency for reading data, consumes additional area resources, and may potentially limit the usefulness of this storage unit for other applications. Despite this, there are several benefits to taking the SPU approach: 
\begin{enumerate}
	\item Parallelism can be more thoroughly exploited than if the algorithm were implemented by a multi-core processor
	\item Vital resources within the host are conserved, allowing the storage unit to act as an accelerator at scale
	\item A key challenge for storage elements is I/O availability: an SPU does not require any additional I/O
	\item Memory-mapped configuration facilitates the design of a microarchitecture that is simple to implement, configure, and extend
	\item An in-path SPU may be sold as locked IP that may be dropped into other SoC environments
	\item Significant power savings are obtained over in-host processing, which operates at an order of magnitude higher frequency
	\item Scalability of the design is significantly improved over external accelerator designs, which are bottlenecked by memory bandwidth limitations
	\item This design can be scaled to other emerging memory devices that offer similar competitiveness as flash, such as memristors or PCM.
\end{enumerate}

A clear explanation of the model utilized to implement the HTM spatial pooler has been presented, providing insight into NVHTM design limitations and benefits. SP, originally developed with high-level programming in mind features significant branching behavior and unclear parameter limits. Despite these challenges, a simplified hardware model has been presented along with an accompanying system model. 
Power, area, and latency estimates were extracted from each phase of the design process to acquire a baseline for feasibility analysis. NVHTM, paired with SVM for classification, present results comparable to those found within the literature for HTM MNIST (91.98\%).

In conclusion, the large delays characteristic of SSDs (when compared to main memory or cache) mean that the latency added by the NVHTM pipeline is orders of magnitude less than the baseline latency for a standard SSD. This supports the case for employing storage processor units in hardware, as has been discussed in the literature. However, the specialization required by the hardware limits the potential for deploying this to large scale markets. Implementing the SPU functionality on a reconfigurable platform integrated into the SSD would be an attractive option, because it has the potential to improve design scalability. 
This framework can be extended to the temporal memory of the HTM, which is designed to model the inter-pattern association across various time steps. However the challenge to modeling it is the unbounded segment growth. To manage this specialized techniques such as dynamic pruning and time multiplexing have to be employed.

\section*{Acknowledgement}
The authors would like to thank both Subutai Ahmed and Yuwei Cao from Numenta for their valuable insights and members of the NanoComputing Research Lab for the rigorous technical discussions on the SP.

\bibliographystyle{IEEEtran}

\bibliography{htmspu_confpaper}

\end{document}